\documentclass[prx,aps,twocolumn,superscriptaddress,longbibliography]{revtex4-1}

\usepackage{listings}
\usepackage{tikz}
\usepackage{times}
\usepackage{graphicx}
\usepackage{float}
\usepackage{latexsym,amsmath,amssymb,bm,euscript}
\usepackage{color}
\usepackage{subfigure}
\usepackage{epstopdf}
\usepackage[colorlinks=true,linkcolor=blue,citecolor=blue]{hyperref}
\usepackage{soul}
\usepackage[normalem]{ulem}
\usepackage{mathrsfs}
\usepackage{amssymb}
\usepackage{algpseudocode}
\usepackage{algorithm}
\usepackage{amsmath}
\usepackage{braket}
\usepackage{booktabs}
\usepackage{chngcntr}
\usepackage{lettrine}
\usepackage{bbding}
\usepackage{xspace}
\usepackage{textcomp}
\usepackage{textcase}
\usepackage{setspace}
\usepackage[T1]{fontenc}
\usepackage{float}
\usepackage{graphicx}
\usepackage{multirow}
\usepackage{mathtools}
\usepackage{threeparttable}
\usepackage[vlined, ruled, algo2e, linesnumbered]{algorithm2e}
\usepackage{changes}
\usepackage{makecell}
\usepackage[title]{appendix}

\usetikzlibrary{shapes}

\begin{document}

\title{Structural, magnetic and magnetocaloric properties of triangular-lattice transition-metal phosphates}

\author{Chuandi Zhang}
\thanks{These authors contributed equally to this work.}
\affiliation{School of Physics, Beihang University, Beijing 100191, China}

\author{Junsen Xiang}
\thanks{These authors contributed equally to this work.}
\affiliation{Beijing National Laboratory for Condensed Matter Physics, Institute of Physics, Chinese Academy of Sciences, 100190 Beijing, China}

\author{Quanliang Zhu}
\affiliation{School of Physics, Beihang University, Beijing 100191, China}

\author{Longfei Wu}
\affiliation{School of Physics, Beihang University, Beijing 100191, China} 

\author{Shanfeng Zhang}
\affiliation{School of Physics, Beihang University, Beijing 100191, China} 

\author{Juping Xu}
\affiliation{Spallation Neutron Source Science Center, Dongguan 523803, China} 

\author{Wen Yin}
\affiliation{Spallation Neutron Source Science Center, Dongguan 523803, China}

\author{Peijie Sun}
\affiliation{Beijing National Laboratory for Condensed Matter Physics, Institute of Physics, Chinese Academy of Sciences, 100190 Beijing, China}

\author{Wei Li}
\affiliation{CAS Key Laboratory of Theoretical Physics, Institute of Theoretical Physics, Chinese Academy of Sciences, Beijing 100190, China}
\affiliation{CAS Center for Excellence in Topological Quantum Computation, University of Chinese Academy of Sciences, Beijng 100190, China}
\affiliation{Peng Huanwu Collaborative Center for Research and Education, Beihang University, Beijing 100191, China}

\author{Gang Su}
\affiliation{Kavli Institute for Theoretical Sciences, and 
		School of Physical Sciences, University of Chinese Academy of Sciences, Beijng 100049, China}
\affiliation{CAS Center for Excellence in Topological 
		Quantum Computation, University of Chinese Academy 
		of Sciences, Beijng 100190, China}

\author{Wentao Jin}
\email{wtjin@buaa.edu.cn}
\affiliation{School of Physics, Beihang University, Beijing 100191, China}

\date{\today}

\begin{abstract}
The recent discovery of the spin supersolid candidate Na$_2$BaCo(PO$_4$)$_2$ stimulates numerous research interest on the triangular-lattice transition-metal phosphates. Here we report a comprehensive study on the structural, magnetic and magnetocaloric properties of polycrystalline Na$_2$$A$$T$(PO$_4$)$_2$ ($A$ = Ba, Sr; $T$ = Co, Ni, Mn). 
  X-ray and neutron diffraction measurements confirm that Na$_2$Ba$T$(PO$_4$)$_2$ (NB$T$P) crystallizes in a trigonal structure, while 
  Na$_2$Sr$T$(PO$_4$)$_2$ (NS$T$P) forms a monoclinic structure with a slight distortion of the triangular network of $T^{2+}$ ions.  
  The dc magnetization data show that all six compounds order antiferromagnetically below 2~K, and the N\'{e}el temperatures of NS$T$P are consistently higher than those of NB$T$P for $T$ = Co, Ni, and Mn, due to the release of geometrical frustration by monoclinical distortions.  
  Further magnetocaloric measurements show that trigonal NB$T$P can reach a lower temperature in the quasi-adiabatic demagnetization process and thus shows a better performance in the magnetic refrigeration, compared with monoclinic NS$T$P. 
  Our findings highlight the outstanding magnetocaloric performances of the trigonal transition-metal phosphates, and disclose two necessary ingredients for a superior magnetic coolant that can reach an ultra-low temperature, including a perfect geometrically frustrated lattice and a small effective spin number associated with the magnetic ions.

\end{abstract}

\maketitle

\section{Introduction}

Competing magnetic interactions in geometrically frustrated magnets can suppress conventional magnetic order down to very low temperature and sometimes lead to exotic spin states, such as spin ice \cite{Bramwell2001,Castelnovo2011,Gingras2014}, quantum Ising spins \cite{Shen2019,Li2020a,Hu2020,Dun2021}, or quantum spin liquid (QSL) \cite{Balents2010,Zhou2017,Broholm2020,Wen2019}. 
The quasi two-dimensional (2D) triangular-lattice antiferromagnets, as the geometrically simplest frustrated motif, have attracted tremendous research interest in the past two decades \cite{Gaulin1994,Collins1997, LYS2020},  including the well studied QSL candidates $\kappa$-(BEDT-TTF)$_2$Cu$_2$(CN)$_3$ \cite{Shimizu2003,Shimizu2006,Yamashita09}, YbMgGaO$_4$ \cite{LYS2015,Li2015,Shen2016,Paddison2017}, and NaYb$X_2$ ($X$ = O, S, Se) \cite{Liu2018,Ding2019,Sarkar2019,Baenitz2018,Ranjith2019,Dai2020}.

Very recently, Liu $et$ $al.$ \cite{Liu2022} propose that the spin frustrations in the QSL candidates make them superior sub-Kelvin magnetocaloric materials that exhibit prominent cooling effects through the adiabatic demagnetization, especially near the quantum critical points. 
This has opened up many possibilities for the realistic applications of numerous geometrically frustrated magnets, in deep-space explorations or quantum computations, for example. 
So far, hydrated paramagnetic salts such as FeNH$_4$(SO$_4$)$_2$ $\cdot$ 12H$_2$O (FAA) and CrK(SO$_4$)$_2$ $\cdot$ 12H$_2$O (CPA) have been used as the main refrigerants for sub-Kelvin magnetic cooling \cite{Shirron2014,Zheng2016}. 
However, these hydrate paramagnetic salts have multiple disadvantages in their applications. 
On one hand, they are structurally less stable, as the water molecules in these hydrates may get lost if used or kept improperly, which largely limits their usages in ultra-high vacuum environment. 
On the other hand, the low spin density and weak paramagnetic interactions between the magnetic ions significantly constrain the volumetric cooling capacity of them.
Therefore, it is very important to seek for novel types of sub-Kelvin magnetic coolants that are structurally more stable and magnetically denser towards better applications.

The transition-metal phosphates Na$_2$$A$$T$(PO$_4$)$_2$ ($A$ = Ba, Sr; $T$ = Co, Ni, Mn) with the glaserite-type structure, in which magnetic $T^{2+}$ ions form a perfect or slightly deformed triangular network, have attracted much attention in the past few years. 
As a quasi 2D antiferromagnet with an effective spin of $J\rm_{eff}$ = 1/2 due to spin-orbit coupling and crystal field splitting, Na$_2$BaCo(PO$_4$)$_2$ (NBCP) was first proposed as a QSL candidate \cite{Zhong2019,Lee2021} but later confirmed to be antiferromagnetically ordered below $\sim$ 150 mK with intriguing spin supersolidity \cite{Li2020,Gao2022a,Sheng2022,Xiang2024}. For comparison, 
Na$_2$BaNi(PO$_4$)$_2$ (NBNP) with $S$ = 1 was reported to order antiferromagnetically below $\sim$ 430 mK \cite{Li2021}, while Na$_2$BaMn(PO$_4$)$_2$ (NBMP) with $S$ = 5/2 was found to display two magnetic transitions at 1.15 and 1.30 K, respectively \cite{Kim2022}. 
The relatively low ordering temperatures, strong spin fluctuations, and abundant field-induced quantum spin state transitions in these equilateral triangular-lattice antiferromagnets render them as promising candidates of sub-Kelvin magnetic refrigerants \cite{Liu2022}. 
On the other hand, it was revealed that the ordering temperature can be lifted by a factor of four by the structural deformation from an equilateral triangular lattice in NBCP to an isoscele triangular lattice in Na$_2$SrCo(PO$_4$)$_2$ (NSCP), due to the release of spin frustration associated with the monoclinic distortion \cite{Zhang2022,Bader2022}. Although the structural and magnetic properties of NBCP, NBNP, NBMP and NSCP have been explored to different extents, 
systematic comparative studies on the trigonal Na$_2$Ba$T$(PO$_4$)$_2$ (NB$T$P) and monoclinic Na$_2$Sr$T$(PO$_4$)$_2$ (NS$T$P) are still missing so far, which will be important for understanding the relationship between the structural, magnetic and magnetocaloric properties of these transition-metal phosphates.

In this study, we have conducted comprehensive structural, magnetic and magnetocaloric studies on polycrystalline samples of the triangular-lattice transition-metal phosphates Na$_2$$A$$T$(PO$_4$)$_2$ ($A$ = Ba, Sr; $T$ = Co, Ni, Mn). Among them, Na$_2$SrNi(PO$_4$)$_2$ (NSNP) and Na$_2$SrMn(PO$_4$)$_2$ (NSMP) were synthesized and characterized for the first time.
It is found that NB$T$P or NS$T$P crystallize in the trigonal or monoclinic structure, in which the magnetic ions form equilateral or isosceles triangular lattices, respectively.  
All six compounds order antiferromagnetically below 2~K, and the N\'{e}el temperatures of NS$T$P are consistently higher than those of NB$T$P for $T$ = Co, Ni, and Mn, as the result of less geometrical frustration due to monoclinical distortions.  By performing the magnetocaloric measurements in the quasi-adiabatic demagnetization, trigonal NB$T$P is found to display a better cooling performance in the magnetic refrigeration over the monoclinic NS$T$P. 
These results reveal two important guidances for finding a superior sub-Kelvin magnetic coolant to attain an ultra-low temperature in the family of frustrated magnets, i.e., a perfect geometrically frustrated lattice and a small effective spin number associated with the magnetic ions.

\section{Methods}

Polycrystalline samples of Na$_2$$AT$(PO$_4$)$_2$ ($A$ = Ba, Sr; $T$ = Co, Ni, Mn) were synthesized by standard solid-state reaction method.  Stoichiometric amounts of dried Na$_2$CO$_3$(99.99$\%$), BaCO$_3$ (99.95$\%$), SrCO$_3$ (99.95$\%$), CoO (99.99$\%$), NiO (99.99$\%$), MnO (99.5$\%$) and (NH$_4$)$_2$HPO$_4$ (99.99$\%$) were mixed and well ground with the catalyst NH$_4$Cl (99.99$\%$) in the molar ratio of 2:1. The pellets were put into alumina crucibles, sintered in air at 800-850 $^\circ$C for 24~h, and cooled down to room temperature. The sintering process was repeated multiple times to minimize possible impurity phases.

The crystal structures of NBCP and NSCP have been thoroughly investigated in our previous study \cite{Zhang2022}, while the structural characterizations of the other four compounds in this study were performed by complementary x-ray diffraction (XRD) and neutron powder diffraction (NPD) measurements at room temperature. 
The XRD patterns were collected using a Bruker D8 ADVANCE diffractometer in Bragg-Brentano geometry with Cu-K$\alpha$  radiation ($\lambda$  = 1.5406 \AA) in the range of 10-90$^\circ$. 
The NPD experiments were conducted on the Multi-Physics Instrument (MPI), a total scattering neutron time-of-flight diffractometer at China Spallation Neutron Source (CSNS), Dongguan, China \cite{xuMultiphysicsInstrumentTotal2021}. 
Polycrystalline samples with the mass of $\sim$ 2.5 grams were loaded into cylindrical containers made of null-scattering TiZr alloy, and the neutron diffraction patterns were collected at room temperature. 
The program FullProf \cite{Rodr1993} was used for the Rietveld refinement of the XRD and NPD patterns and the determination of crystal structures for all compounds.

DC magnetization of all polycrystalline samples were measured using a Quantum Design Magnetic Property Measurement System (MPMS) equipped with the $^{3}$He insert. 
The magnetocaloric effects (MCE) were measured by recording the quasi-adiabatic temperature changes of the sample during the demagnetization process, utilizing a field-calibrated RuO$_2$ thermometer attached onto the surface of the pelletized sample. A homemade sample stage was integrated into the Quantum Design Physical Property Measurement System (PPMS) \cite{xiang2023}, on which the pellet of Na$_2$$AT$(PO$_4$)$_2$ with a total mass of $\sim$ 0.7~g was loaded. To improve the low-temperature thermal conductivity of Na$_2$$AT$(PO$_4$)$_2$ samples in the MCE measurements, silver powders with the mass of $\sim$ 0.35~g was homogeneously mixed into the pellets. An additional guard stage composed of single crystals ($\sim$ 20 g) of Ga$_5$Gd$_3$O$_{12}$ (GGG), one of the most widely used commercial magnetic coolants, serves as the thermal intercept between the sample stage and the PPMS chamber. The MCE measurements of GGG itself with the same guard stage were conducted in the same demagnetization conditions for comparison, using polycrystalline samples of GGG ($\sim$ 1.5 g) and silver powders with the same mass homogeneously mixed into the pellet.

\section{Results and Discussions}

\begin{figure*}[htbp]
  \includegraphics[width = 15.6cm]{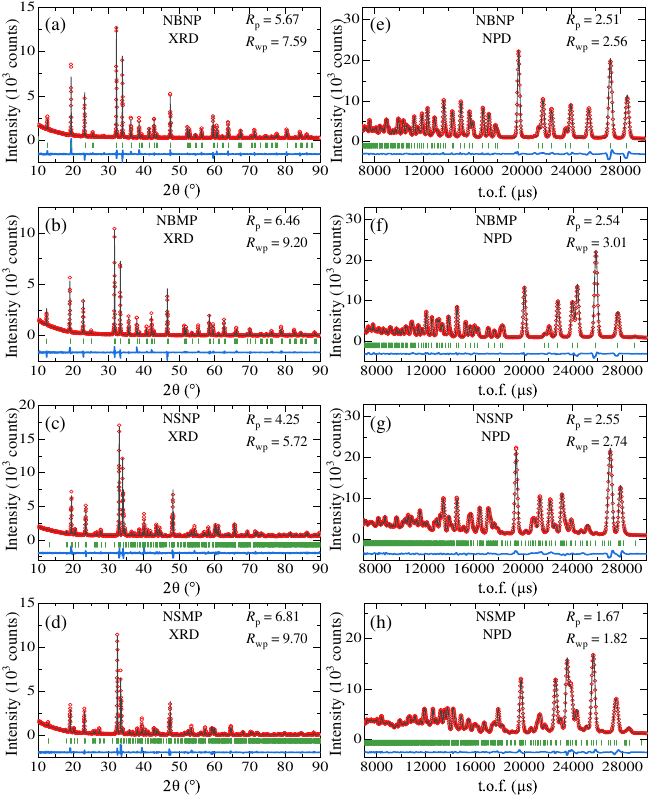}
  \caption{Room-temperature XRD and NPD patterns of NBNP (a, e), NBMP (b, f), NSNP (c, g) and NSMP (d, h) and corresponding Rietveld refinements. The circles represent the observed intensities, and the solid lines are the calculated patterns. The differences between the observed and calculated intensities are shown at the bottoms. The short vertical bars correspond to the expected nuclear Bragg reflections.
}
\label{Fig1}
    \end{figure*}
\vspace{1em}
\noindent

The XRD and NPD patterns of polycrystalline NBNP, NBMP, NSNP and NSMP collected at room temperature are simultaneously refined for structural determinations, as shown in Fig.~\ref{Fig1}. Our previous study has confirmed the trigonal symmetry (space group $\mathit{P}$$\bar{3}$$\mathit{m}$1) of NBCP and the monoclinic symmetry (space group $\mathit{P}$2$_{1}$/$\mathit{a}$) of NSCP \cite{Zhang2022}. 
Similarly, here the new results on the other four compounds indicate trigonal structures of NBNP (Fig.~\ref{Fig1}\textbf{a, e}) and NBMP (Fig.~\ref{Fig1}\textbf{b, f}), in contrast to monoclinic structures of NSNP (Fig.~\ref{Fig1}\textbf{c, g}) and NSMP (Fig.~\ref{Fig1}\textbf{d, h}).

Specifically, it is found that at room temperature the space group is $\mathit{P}$$\bar{3}$ for both NBNP and NBMP. Different from NBCP with the space group of $\mathit{P}$$\bar{3}$$\mathit{m}$1, half of the oxygen atoms in NBNP and NBMP move off the ($x, 1-x$) positions in the $ab$ plane, causing opposite rotations of the NiO$\rm_6$ or MnO$\rm_6$ octahedra and PO$\rm_4$ tetrahedra along the $c$-axis and thus eliminating the mirror symmetry. This can be validated by excellent agreements between the observed and calculated NPD intensities adopting the $\mathit{P}$$\bar{3}$ space group, as shown in Fig.~\ref{Fig1}\textbf{e} and \textbf{f}, since neutrons are very sensitive to the atomic positions of oxygen.
The threefold rotational symmetry is retained, however, with the Ni$^{2+}$ or Mn$^{2+}$ residing on a perfect equilateral triangular network. The result obtained here for NBNP is well consistent with a single-crystal XRD determination, which also reports a space group of $\mathit{P}$$\bar{3}$ \cite{Li2021}. 
However, it is worth noting that the structure of NBMP is controversial so far, as both $\mathit{P}$$\bar{3}$ and $\mathit{P}$$\bar{3}$$\mathit{m}$1 were ever proposed as the possible space group \cite{Nishio-Hamane2014,Nenert2020,Kim2022}.

\begin{figure}[t]
  \includegraphics[width = 8.5cm]{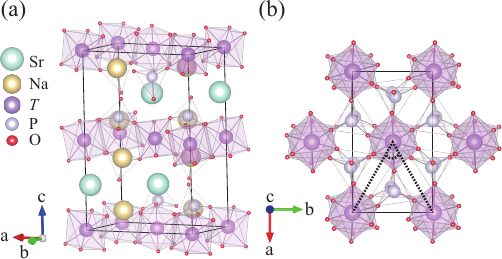}
  \caption{The monoclinic unit cell of NS$T$P (a) with the in-plane structure viewed along the $c$ axis illustrated in (b), where the angle enclosed by the dashed line marks the in-plane $T$-$T$-$T$ bond angle in the isosceles triangular network of the $T^{2+}$ layers.}
 \label{Fig2}
    \end{figure}

On the other hand, a lattice deformation is clearly featured when Ba is substituted by Sr. Fig.~\ref{Fig2}\textbf{a} shows the monoclinic unit cell of NS$T$P as determined by the structural refinements. To compensate the considerable difference of the radius between Sr$^{2+}$ and Ba$^{2+}$ ions, a structural transformation takes place due to the rotations of $T$O$_6$ octahedra, the distortion of PO$_4$ tetrahedra and the displacement of Sr$^{2+}$ ions, which breaks the threefold rotational symmetry and deforms the network of magnetic $T^{2+}$ ions into an isosceles triangular lattice (see Fig.~\ref{Fig2}\textbf{b}). As a result, the in-plane $T$-$T$-$T$ bond angle in the isosceles triangles of the $T^{2+}$ layers, as listed in Table.~\ref{Table1},  deviates from 60$^\circ$ for NSCP, NSNP, and NSMP. And the degree of monoclinic distortion is comparable for the NS$T$P compounds. As shown in Fig.~\ref{Fig1}\textbf{c, d, g, h}, the XRD and NPD diffraction patterns of either NSNP or NSMP can be nicely fitted by a monoclinic structure with the same space group of $\mathit{P}$2$_{1}$/$\mathit{a}$ as NSCP.

\begin{table*}[]
  \caption{\textbf{}The space group and refined lattice constants of Na$_2$$A$$T$(PO$_4$)$_2$ ($A$ = Ba, Sr; $T$ = Co, Ni, Mn) at room temperature.}
  \begin{ruledtabular}
		\resizebox{\linewidth }{!}{  
  \begin{tabular}{ccccccc}
    ~~~~Compounds ~~~ & ~~~~Space group ~~~ &  ~~~~$a$ (\AA) ~~~ &  ~~~~$b$ (\AA) ~~~ &  ~~~~$c$ (\AA) ~~~ &  ~~~~$\beta$ ($^\circ$) ~~~ &  ~~~~in-plane $T$-$T$-$T$ bond angle ($^\circ$) ~~~ \\ 
  \hline
  NBCP\cite{Zhang2022}      & $\mathit{P}$$\bar{3}$$\mathit{m}$1          & 5.313(1)                 & 5.313(1)                 & 7.007(1)                 & 90  &  60                  \\
  NBNP      & $\mathit{P}$$\bar{3}$          & 5.285(1)                 & 5.285(1)                 & 6.985(1)                 & 90       &      60          \\
  NBMP      & $\mathit{P}$$\bar{3}$          & 5.373(1)                 & 5.373(1)                 & 7.097(1)                 & 90     &       60           \\
  NSCP\cite{Zhang2022}      & $\mathit{P}$2$_{1}$/$\mathit{a}$          & 9.194(1)                 & 5.259(1)                 & 13.531(1)                & 90.07(1)    &     59.54(1)            \\
  NSNP      & $\mathit{P}$2$_{1}$/$\mathit{a}$          & 9.148(1)                 & 5.240(1)                 & 13.495(1)                & 89.98(1)      &  59.61(1)             \\
  NSMP      & $\mathit{P}$2$_{1}$/$\mathit{a}$           & 9.301(1)                 & 5.326(1)                 & 13.699(1)                & 90.09(1)& 59.59(1) \\                  
  \end{tabular}
}
	\end{ruledtabular}
\label{Table1}
  \end{table*}

\begin{table}[ht]
  \caption{\textbf{}Refined atomic parameters of Na$_2$BaNi(PO$_4$)$_2$ (NBNP) at room temperature.}
\begin{ruledtabular}
		\resizebox{\columnwidth }{!}{  
\begin{tabular}{cccccl}
  Atom & ~~~~Site~~~~ & ~~~~~~$x$~~~~~~ & ~~~~~~$y$~~~~~~ & ~~~~~~$z$~~~~~~ & ~~~~~~$B\rm_{iso}$(\AA$^2$)  \\ \hline
  Ba      &1a   & 0.000   & 0.000   & 0.000     &~~~~~~~~0.36(1)   \\
  Ni      &1b   & 0.000   & 0.000   & 0.500     &~~~~~~~~0.22(1)   \\
  P       &2d   & 0.333   & 0.666   & 0.244(1)      &~~~~~~~~0.11(1)   \\
  Na      &2d   & 0.333   & 0.666   & 0.676(1)      &~~~~~~~~1.36(1)   \\
  O1   &2d   & 0.333   & 0.666   & 0.027(1)      &~~~~~~~~0.69(1)   \\
  O2   &6g   & 0.228(1)   & 0.873(1)   & 0.321(1)      &~~~~~~~~0.69(1)   \\       
  \end{tabular}
}
	\end{ruledtabular}
\label{Table2}
  \end{table}

\begin{table}[h]
  \caption{\textbf{}Refined atomic parameters of Na$_2$BaMn(PO$_4$)$_2$ (NBMP) at room temperature.}
  \begin{ruledtabular}
		\resizebox{\columnwidth }{!}{  
  \begin{tabular}{cccccl}
    Atom & ~~~~Site~~~~ & ~~~~~~$x$~~~~~~ & ~~~~~~$y$~~~~~~ & ~~~~~~$z$~~~~~~ & ~~~~~~$B\rm_{iso}$(\AA$^2$)  \\ \hline
  Ba      &1a   & 0.000   & 0.000   & 0.000      &~~~~~~~~1.34(1)   \\
  Mn      &1b   & 0.000   & 0.000   & 0.500      &~~~~~~~~1.06(1)   \\
  P       &2d   & 0.333   & 0.666   & 0.764(1)      &~~~~~~~~0.76(1)   \\
  Na      &2d   & 0.666   & 0.333   & 0.685(1)      &~~~~~~~~1.47(1)   \\
  O1  &6g   & 0.128(1)   & 0.361(1)   & 0.687(1)      &~~~~~~~~1.04(1)   \\
  O2   &2d   & 0.333   & 0.666   & 0.977(1)     &~~~~~~~~1.04(1)   \\                 
  \end{tabular}
  }
	\end{ruledtabular}
\label{Table3}
  \end{table}

\begin{table}[h]
  \caption{\textbf{}Refined atomic parameters of Na$_2$SrNi(PO$_4$)$_2$ (NSNP) at room temperature.}
  \begin{ruledtabular}
		\resizebox{\columnwidth }{!}{ 
  \begin{tabular}{cccccl}
    Atom & ~~Site~~ &~~$x$~~ & ~~$y$~~ & ~~$z$~~ &~~ $B\rm_{iso}$(\AA$^2$)  \\ \hline
  Na1      &4e   &~~~0.170(1)   &~~~0.511(2)   &~~~0.903(1)      &~~~~~0.24(2)   \\
  Na2      &4e   &~~~ 0.153(1)   &~~~ 0.539(2)   &~~~ 0.419(1)      &~~~~~0.24(2)   \\
  Sr       &4e   &~~~ 0.034(1)   &~~~ 0.035(1)   &~~~ 0.749(1)      &~~~~~0.43(1)   \\
  Ni1      &2d   &~~~ 0.000   &~~~ 0.000    &~~~ 0.500      &~~~~~0.01(2)   \\
  Ni2   &2a   &~~~ 0.000   &~~~ 0.000   &~~~ 0.000      &~~~~~0.01(2)   \\
  P1  &4e   &~~~ 0.187(1)   &~~~ 0.515(1)   &~~~ 0.632(1)      &~~~~~0.03(1)   \\ 
  O1      &4e   &~~~ 0.157(1)   &~~~ 0.513(1)   &~~~ 0.741(1)      &~~~~~0.06(1)   \\
  O2      &4e   &~~~ 0.346(1)   &~~~ 0.499(1)   &~~~ 0.608(1)      &~~~~~0.06(1)   \\
  O3       &4e   &~~~ 0.095(1)   &~~~ 0.295(1)   &~~~ 0.577(1)      &~~~~~0.06(1)   \\
  O4      &4e   &~~~ 0.132(2)   &~~~ 0.779(1)   &~~~ 0.586(1)     &~~~~~0.06(1)   \\
  P2   &4e   &~~~ 0.157(1)   &~~~ 0.494(1)   &~~~ 0.132(1)      &~~~~~0.03(1)   \\
  O5   &4e   &~~~ 0.178(1)   &~~~ 0.542(1)   &~~~ 0.242(1)      &~~~~~0.02(1)   \\ 
  O6      &4e   &~~~ 0.298(1)   &~~~ 0.499(1)   &~~~ 0.081(1)     &~~~~~0.02(1)   \\
  O7      &4e   &~~~ 0.077(1)   &~~~ 0.224(1)   &~~~ 0.112(1)      &~~~~~0.02(1)  \\
  O8       &4e   &~~~ 0.053(1)   &~~~ 0.701(1)   &~~~ 0.092(1)      &~~~~~0.02(1)   \\                  
  \end{tabular}
  }
	\end{ruledtabular}
\label{Table4}
  \end{table}

\begin{table}[h]
  \caption{\textbf{}Refined atomic parameters of Na$_2$SrMn(PO$_4$)$_2$ (NSMP) at room temperature.}
  \begin{ruledtabular}
		\resizebox{\columnwidth }{!}{ 
  \begin{tabular}{cccccl}
    Atom & ~~Site~~ &~~$x$~~ & ~~$y$~~ & ~~$z$~~ &~~ $B\rm_{iso}$(\AA$^2$)  \\ \hline
  Na1      &4e   &~~~ 0.180(1)   &~~~ 0.519(2)   &~~~ 0.906(1)     &~~~~~0.09(2)   \\
  Na2      &4e   &~~~ 0.157(1)   &~~~ 0.521(2)   &~~~ 0.410(1)      &~~~~~0.09(2)   \\
  Sr       &4e   &~~~ 0.034(1)   &~~~ 0.039(1)   &~~~ 0.747(1)      &~~~~~0.07(1)   \\
  Mn1      &2d   &~~~ 0.000   &~~~ 0.000    &~~~ 0.500      &~~~~~0.01(2)   \\
  Mn2   &2a   &~~~ 0.000   &~~~ 0.000   &~~~ 0.000     &~~~~~0.01(2)   \\
  P1   &4e   &~~~ 0.175(1)   &~~~ 0.521(1)   &~~~ 0.637(1)     &~~~~~0.03(1)   \\ 
  O1      &4e   &~~~ 0.141(1)   &~~~ 0.509(1)   &~~~ 0.743(1)      &~~~~~0.09(1)   \\
  O2      &4e   &~~~ 0.343(1)   &~~~ 0.508(1)   &~~~ 0.616(1)     &~~~~~0.09(1)   \\
  O3       &4e   &~~~ 0.102(1)   &~~~ 0.301(1)   &~~~ 0.584(1)      &~~~~~0.09(1)   \\
  O4     &4e   &~~~ 0.123(1)   &~~~ 0.780(1)   &~~~ 0.594(1)      &~~~~~0.09(1)   \\
  P2   &4e   &~~~ 0.153(1)   &~~~ 0.491(1)   &~~~ 0.136(1)      &~~~~~0.03(1)   \\
  O5   &4e   &~~~ 0.182(1)   &~~~ 0.545(1)   &~~~ 0.242(1)      &~~~~~0.01(1)   \\ 
  O6     &4e   &~~~ 0.294(1)   &~~~ 0.497(1)   &~~~ 0.076(1)     &~~~~~0.01(1)   \\
  O7      &4e   &~~~ 0.087(1)   &~~~ 0.235(1)   &~~~ 0.118(1)     &~~~~~0.01(1)  \\
  O8       &4e   &~~~ 0.048(1)   &~~~ 0.683(1)   &~~~ 0.093(1)      &~~~~~0.01(1)   \\              
  \end{tabular}
  }
	\end{ruledtabular}
\label{Table5}
  \end{table}

In Table.~\ref{Table1}, we have listed the lattice constants of all six compounds in this study for comparison, which are well consistent with the difference between the effective ionic radius of the $T^{2+}$ ions (0.74 \AA $ $ for Co$^{2+}$, 0.69 \AA $ $ for Ni$^{2+}$ and 0.83 \AA $ $ for Mn$^{2+}$). In addition, all atomic parameters in NBNP, NBMP, NSNP, and NSMP, as determined by the Rietveld refinements in the current study, are listed in Table.~\ref{Table2}, ~\ref{Table3}, ~\ref{Table4}, ~\ref{Table5}, respectively.

\begin{figure*}[ht]
  \includegraphics[width = 17.6cm]{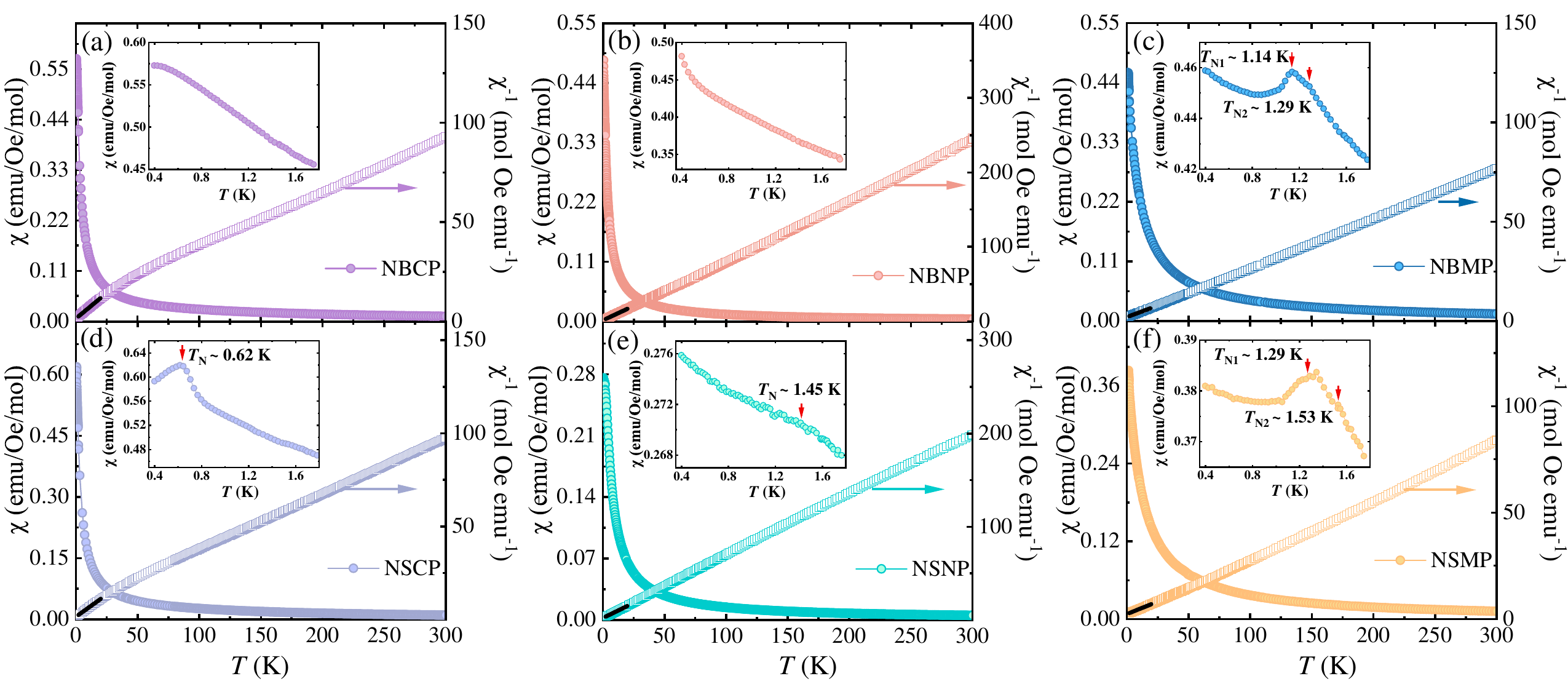}
  \caption{DC magnetic susceptibility ($\chi$, filled circles) and inverse susceptibility (1/$\chi$, empty squares) of the polycrystalline NBCP (a), NBNP (b), NBMP (c), NSCP (d), NSNP (e) and NSMP (f), respectively, measured in a magnetic field of 0.1~T.  The black solid lines represent the CW fittings to 1/$\chi$ in the low-temperature paramagnetic state from 2 to 20 K. The insets are enlarged plots of $\chi(T)$ in the temperature range from 0.4 to 1.8 K, in which the arrows mark the AFM ordering temperatures.}
\label{Fig3}
    \end{figure*}

The degree of geometrical frustration in equilateral and isoscele triangular lattices differs, which inspires us to study the relationship between their structural and magnetic properties. 
Fig.~\ref{Fig3} shows the dc magnetic susceptibility ($\chi$) of polycrystalline Na$_2$$A$$T$(PO$_4$)$_2$ as a function of temperature from 0.4 K to 300 K, measured in an applied magnetic field of 0.1 T. None of them order magnetically above 2 K. By doing the Curie-Weiss (CW) fittings to the inverse susceptibility (1/$\chi$) from 2 K to 20 K in the low-temperature paramagnetic state to exclude the influence of crystal field excitations, as shown in the Fig.~\ref{Fig3}, the CW temperature ($\theta\rm_{cw}$) and effective moment ($\mu\rm_{eff}$) of the $T^{2+}$ ions can be estimated. As shown in Table.~\ref{Table6}, negative values of $\theta\rm_{cw}$ for all six compounds suggest dominant antiferromagnetic (AFM) interactions within the triangular-lattice layers, and the extracted values of $\mu\rm_{eff}$ for polycrystalline NBCP, NBNP, NBMP and NSCP are in good agreement with those estimated for single-crystal samples \cite{Li2020,Li2021,Kim2022,Bader2022}. In addition, according to $\mu$$\rm _{eff}$ = $g$$\sqrt{S(S+1)}$$\mu$$\rm _{B}$, the Land\'{e} factors ($g$) of the polycrystalline samples can also be calculated and listed in Table.~\ref{Table6}.

\begin{table*}[]
  \caption{\textbf{}Magnetic parameters of Na$_2$$A$$T$(PO$_4$)$_2$ ($A$ = Ba, Sr; $T$ = Co, Ni, Mn).} 
  \begin{ruledtabular}
		\resizebox{\linewidth}{!}{ 
  \begin{tabular}{ccccccc}
    ~~~~Sample ~~~ & ~~~~effective spin number ~~~ &  ~~~ g factor ~~~ &  ~~~~$H$$\rm _s$ (T)  at 0.4 K~~~ &  ~~~~$\theta_{cw}$ (K)  ~~~ &  ~~~~$\mu$$\rm _{eff}$ ($\mu$$\rm _B$)   ~~~& $T$$\rm _N$ (K)  \\ \hline
  NBCP            & 1/2    & 4.57(1)  & 2.61(2)      & $-$2.85(11)      & 3.96(1) & $\sim$ 150~mK \cite{Li2020,Sheng2022,Xiang2024} \\ \hline
  NBNP            & 1   & 2.28(1)   & 3.66(2)      & $-$1.99(2)       & 3.22(1)   & $\sim$ 430~mK \cite{Li2021}   \\ \hline
  NBMP            & 5/2  & 1.98(1)   & no saturation observed      & $-$8.39(14)      & 5.86(1)   & \makecell{1.14~K, 1.29~K (this work)  \\ 1.15 K, 1.30 K \cite{Kim2022}} \\ \hline
  NSCP            & 1/2  & 4.73(1)    & 3.15(2)      & $-$2.98(13)      & 4.10(1)   &\makecell{620~mK (this work)     \\ $\sim$ 600 mK \cite{Bader2022}}  \\ \hline
  NSNP           & 1  & 2.47(1)    & no saturation observed      & $-$2.97(15)      & 3.50(2)   & $\sim$ 1.45~K (this work)  \\ \hline
  NSMP           & 5/2  & 1.98(1)    & no saturation observed      & $-$9.90(12)      & 5.85(1)    & 1.29~K, 1.53~K (this work) \\                   
  \end{tabular}
  }
\end{ruledtabular}
\label{Table6}
  \end{table*}

By inspecting $\chi(T)$ at very low temperature from 0.4 to 1.8 K, it is found that trigonal NBCP shows a smooth increase of $\chi$ upon cooling and absence of magnetic ordering down to 0.4 K (see the inset of Fig.~\ref{Fig3}\textbf{a}), which is consistent with previous reports on the occurrence of AFM ordering only below $T\rm_{N}\simeq$ 150 mK \cite{Li2020,Sheng2022,Xiang2024}. 
In contrast, monoclinic NSCP displays an AFM-like transition at a much higher temperature of $\sim$ 620 mK (see Fig.~\ref{Fig3}\textbf{d}), similar to $T\rm_{N}\simeq$ 600 mK as reported for single-crystal NSCP \cite{Bader2022}. This indicates a much weaker frustration effect in the isosceles triangular lattice of Co associated with the monoclinic distortion, compared with NBCP with a perfect equilateral triangular lattice of Co. In addition, trigonal NBNP with $S$ = 1 begins to rise sharply in $\chi$ around 450 mK (see Fig.~\ref{Fig3}\textbf{b}), which is indicative of the onset of AFM ordering and agrees well with $T\rm_{N}\simeq$ 430 mK for single-crystal NBNP \cite{Li2021}. Trigonal NBMP with $S$ = 5/2, however, shows two features in $\chi$ at 1.14 K and 1.29 K (see Fig.~\ref{Fig3}\textbf{c}), respectively, again in excellent agreement with the report that the NBMP single crystal exhibits two magnetic transitions at 1.15 K and 1.30 K \cite{Kim2022}. 
To the best of our knowledge, no reports on monoclinic NSNP and NSMP are available yet for comparison. In our study, both NSNP and NSMP, which were synthesized for the first time, are found to order at higher temperatures compared with the trigonal NBNP and NBMP. A broad hump in $\chi$ at $\sim$ 1.45 K is observed for NSNP (see Fig.~\ref{Fig3}\textbf{e}), and two anomalies at 1.29 K and 1.53 K (see Fig.~\ref{Fig3}\textbf{f}), respectively, are revealed for NSMP. 
Although the natures of the magnetic transitions shown in Fig.~\ref{Fig3} are not completely understood for these compounds, a conclusion can be drawn through the comparisons. 
The N\'{e}el temperatures of NS$T$P are consistently higher than those of corresponding NB$T$P for $T$ = Co, Ni, and Mn, due to the release of geometrical frustration by monoclinical distortions. 

\begin{figure*}[htbp]
  \includegraphics[width = 17.6cm]{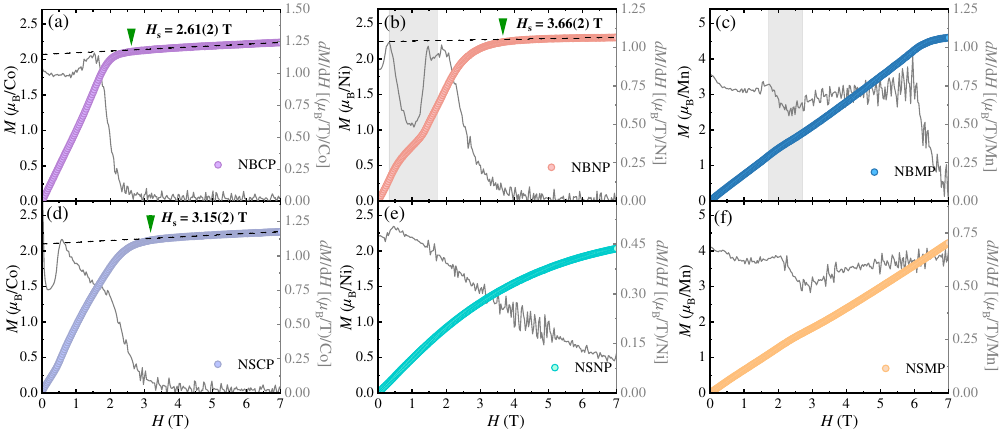}
  \caption{Isothermal magnetization ($M(H)$, filled circles) of polycrystalline NBCP (a), NBNP (b), NBMP (c), NSCP (d), NSNP (e) and NSMP (f), respectively, measured at 0.4 K and their first derivatives ($dM/dH$, solid lines). 
  Dashed lines in (a), (b) and (d) correspond to the contribution from the Van-Vleck paramagnetism.
    The shaded zones in (b) and (c) mark the region of the slope change of the $M(H)$ curve, for which a dip appears in $dM/dH$ for NBNP and NBMP. The green arrows in (a), (b) and (d) mark the critical field ($H\rm_{s}$) corresponding to the moment saturation for NBCP, NBNP and NBMP, respectively.
  }
  \label{Fig4}
    \end{figure*}
Furthermore, the isothermal magnetization curves of all six compounds were measured at 0.4 K. As shown in Fig.~\ref{Fig4}, after subtracting the contributions from Van-Vleck paramagnetism, the saturation moment of the magnetic ions is estimated to be 2.07, 2.10 and 2.24~$\mu \rm_B$ for NBCP, NSCP and NBNP, respectively. 
The critical field ($H\rm_s$) at 0.4 K corresponding to the moment saturation is determined to be 2.61(2), 3.15(2), and 3.66(2) T, for NBCP, NSCP and NBNP, respectively. The values of $H\rm_s$ are consistent with the previous studies on these three compounds with single-crystal samples \cite{Li2020,Li2021,Kim2022}, considering the powder averaging effect and higher measurement temperature. For NBMP (Fig.~\ref{Fig4}\textbf{c}), NSNP (Fig.~\ref{Fig4}\textbf{e}) and NSMP (Fig.~\ref{Fig4}\textbf{f}), the moment saturation is not reached up to 7 T and a larger magnetic field is needed to fully polarize the system. Therefore, it is clear that the spin interactions strengthen with increasing effective spin number.

More importantly, as shown in Fig.~\ref{Fig4}\textbf{b, c}, NBNP and NBMP with the trigonal symmetry both exhibit a slope change of the $M(H)$ curve at 0.4 K as a remnant of the zero-temperature one-third magnetization plateau melt by thermal fluctuations, for 0.33 T $\leq$ $H$ $\leq$ 1.74 T and 1.71 T $\leq$ $H$ $\leq$ 2.73 T, respectively, evidenced by a dip in the derivative of the isothermal magnetization ($dM/dH$), which is expected for an "up-up-down" (UUD) quantum spin state preserving the U(1) spin rotational symmetry. 
We notice that the observation of such a one-third magnetization plateau or slope change was well documented for NBCP, NBNP and NBMP single crystals with a field applied along the $c$ axis \cite{Li2020,Li2021,Kim2022}, but here even for polycrystalline NBNP and NBMP we were still able to observe some imprints of this intriguing quantum spin state below the AFM ordering temperature despite of the powder averaging. 
The reason for the absence of one-third plateau in Fig.~\ref{Fig4}\textbf{a} for NBCP is that the measurement temperature of 0.4 K is already beyond the UUD phase regime of this compound \cite{Xiang2024}. On the other hand, the one-third plateau can not be clearly identified for the monoclinic compounds as shown in Fig.~\ref{Fig4}\textbf{d}-\textbf{f}, indicating the importance of a perfect equilateral triangular lattice for the realization of this intriguing field-induced UUD phase. The nature of the seemingly kink around 2.5 T for NSMP in Fig.~\ref{Fig4}\textbf{f} needs to be understood, based on more measurements on single-crystal samples, when available.


\begin{figure}[t]
  \includegraphics[width = 8.5cm]{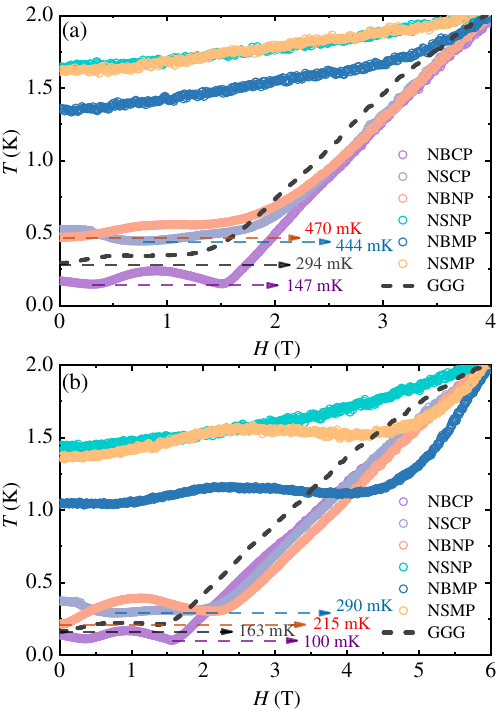}
  \caption{The cooling curve of polycrystalline Na$_2$$A$$T$(PO$_4$)$_2$ samples in the adiabatic demagnetization process starting from an initial temperature of $T\rm_0$ = 2 K and an initial magnetic field of $H\rm_0$ = 4~T (a) and 6~T (b), respectively. The cooling performances of the commercial sub-Kelvin magnetic coolant GGG in similar conditions are also shown, in dashed lines, for comparison.}
 \label{Fig5}
    \end{figure}


Inspired by the recently reported giant magnetocaloric effect (MCE) in NBCP \cite{Xiang2024}, we have systematically tested the cooling performances of all six polycrystalline samples in the quasi-adiabatic demagnetization process. 
As shown in Fig.~\ref{Fig5}\textbf{a,b}, starting from the initial conditions of $T\rm_0$ = 2 K and $H\rm_0$ = 4 T or 6 T realized by PPMS, the demagnetization cooling curve of polycrystalline NBCP shows two prominent valley-like features and strikingly reaches the lowest temperature of 147 mK for $H\rm_0$ = 4 T and 100 mK for $H\rm_0$ = 6 T, respectively.  The magnetic fields corresponding to the two valleys, $\sim$ 0.31(2) T and $\sim$ 1.55(2) T, agree well with the critical field of the spin state changing from a Y to an UUD state for $H$ // $c$ and from a $\tilde{\rm{V}}$ state to the polarized state for $H$ // $a$, respectively, as assigned by the many-body calculations \cite{Gao2022a}. This clearly suggests that the giant MCE observed for NBCP mainly originates from the strong spin fluctuations associated with the quantum criticalities. 
Similarly, polycrystalline NBNP and NSCP also display multiple valley-like features associated with field-induced quantum phase transitions and can reach the lowest temperatures of 215 mK and 290 mK around 0.01 T and 0.77 T, respectively, in the demagnetization process with $H\rm_0$ = 6 T. To better understand the nature of spin state transitions in these compounds, in-depth experimental studies on high-quality single-crystal samples are desired. In the previous MCE study of NBCP single crystals, NBCP clearly exhibits three spin state transitions in the quasi-adiabatic demagnetization process and reaches down to the lowest temperature of 94 mK starting from the initial conditions of $T\rm_0$ = 2 K and $H\rm_0$ = 4 T, owing to the tremendous transverse spin fluctuations in the spin supersolid states \cite{Xiang2024}.

It is worth pointing out that the lowest temperature achieved using polycrystalline NBCP with an effective mass of $\sim$ 0.7~g, although higher than that achieved using single-crystal samples, still outperforms that achieved with polycrystalline GGG, the widely used commercial magnetic coolants, with an effective mass of $\sim$ 1.5~g. As shown in Fig.~\ref{Fig5}, GGG shows the lowest temperature of 294 mK and 163 mK, in a similar demagnetization process with $H\rm_0$ = 4 T and 6 T, respectively. In addition, despite of a much smaller effective spin number ($J\rm_{eff} = 1/2$) compared with GGG ($J$ = 7/2), NBCP shows a larger magnetic entropy below 400 mK and thus a stronger cooling capacity over GGG at ultra-low temperature (see the comparison of their low-temperature specific heat and magnetic entropy data of single-crystal samples shown in Fig.~\ref{Fig6} in Appendix A).

In contrast, trigonal NBMP, monoclinic NSNP and monoclinic NSMP show much worse cooling performances, only reaching the lowest temperatures of 1.04, 1.44 and 1.36 K, respectively, in the quasi-adiabatic demagnetization process with $H\rm_0$ = 6 T, as shown in Fig.~\ref{Fig5}\textbf{b}. 
Looking back into the magnetic susceptibility data in Fig.~\ref{Fig3}, the dramatically distinct MCE behaviors of these three compounds can be attributed to the much higher AFM ordering temperatures and critical fields compared with NBCP, NBNP and NSCP. As a result, there are no remnant low-temperature magnetic entropies and no spin fluctuations associated with the quantum criticalities below 6 T, which accout for their bad cooling performances at very low temperature.


Comparing the magnetocaloric properties of the six compounds we studied, two conclusions can be drawn. 
First, trigonal NB$T$P with a perfect equilateral triangular lattice shows better cooling performances over monoclinic NS$T$P with an isoscele triangular lattice, due to stronger spin frustration effects and  lower magnetic ordering temperatures. 
Second, among the trigonal NB$T$P or monoclinic NS$T$P family, cobaltates ($J\rm_{eff} = 1/2$) can reach the lowest temperature compared to nickelates ($S = 1$) and manganates ($S = 5/2$) in the same ADR process, suggesting the importance of a small spin number, weaker spin interactions and again lower ordering temperatures for sub-Kelvin magnetic referigeration.

~
\section{Conclusion}

In summary, we have synthesized high-purity polycrystalline samples of Na$_2$$A$$T$(PO$_4$)$_2$ ($A$ = Ba, Sr; $T$ = Co, Ni, Mn), the triangular-lattice transition-metal phosphates. These compounds exhibit distinct structural, magnetic and magnetocaloric properties. Compared with trigonal NB$T$P in which the magnetic $T^{2+}$ ions form a perfect equilateral triangular lattice, monoclinic NS$T$P with an isosceles magnetic triangular network clearly show higher AFM ordering temperatures, due to the release of spin frustration, and worse magnetocaloric performance in the quasi-adiabatic demagnetization. Our work indicates the importance of a perfect geometrical frustrated lattice without structural distortion and a small effective spin number in the magnetic refrigeration down to an ultra low temperature. Compared with the hydrate paramagnetic salts, the traditional sub-Kelvin coolants, the structurally more stable and magnetically denser frustrated magnets hold great potentials for applications in the cryogenic refrigeration.

\begin{acknowledgments}

This work is financially supported by the National Key Projects for Research and Development of China (Grant No. 2023YFA1406003), the National Natural Science Foundation of China (Grant No. 12074024), , the Large Scientific Facility Open Subject of Songshan Lake (Dongguan, Guangdong), and the Fundamental Research Funds for the Central Universities in China.

\end{acknowledgments}

\begin{appendices}
\section{Low-temperature specific heat and magnetic entropy of NBCP and GGG}

\begin{figure}[t]
  \includegraphics[width = 8.5cm]{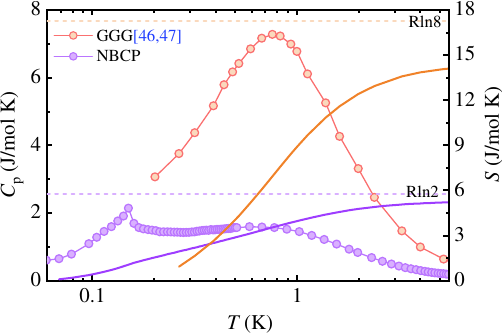}
  \caption{The zero-field molar specific heat ($C$, filled circles) and magnetic entropy ($S$, solid lines) of NBCP and GGG single crystals, respectively, at low temperatures. The specific heat data of single-crystal GGG are from Ref. \onlinecite{onn1967a, schiffer1994}. $S$ is obtained by integrating $C/T$. The horizontal dashed lines mark the maximal magnetic entropy of $R$ln2 and $R$ln8 as expected for the $J\rm_{eff}$ = 1/2 and $J$ = 7/2 systems, respectively.}
 \label{Fig6}
\end{figure}


Fig.~\ref{Fig6} shows the zero-field molar specific heat ($C$) and magnetic entropy ($S$) of NBCP and GGG, based on the low-temperature specific heat measurements on our own single-crystal sample of NBCP and those on single-crystal samples of GGG as reported in previous studies \cite{onn1967a, schiffer1994}, respectively. $S$ is obtained by integrating $C/T$, assuming that the phonon contribution to the specific heat is negligible at such low temperature. For NBCP, a relatively sharp peak around $\sim$ 150 mK in the $C(T)$ curve is consistent with a long-range AFM order into a spin supersolid phase \cite{Gao2022a, Xiang2024}. For GGG, a broad peak in the $C(T)$ curve is indicative of short-range AFM correlations established below $\sim$ 760 mK \cite{schiffer1994}. Because of the much larger spin number of Gd$^{3+}$ ($J$ = 7/2) compared with Co$^{2+}$ ($J\rm_{eff}$ = 1/2), GGG exhibits a dramatically larger maximal magnetic entropy $S\rm_{m}$ = $R$ln(2$J$+1) than NBCP. However, it's noticeable that NBCP shows a larger magnetic entropy below 400 mK and thus a stronger cooling capacity over GGG at ultra-low temperature. 

\section{Estimations about the eddy current heating effect of silver powders}

The primary role of silver powders in the MCE measurements is to reduce the interfacial thermal resistance between the large amounts of grains of the polycrystalline sample. This facilitates the rapid transmission of internal coldness to the surface of the sample pellet, thereby reducing the heat loss and ensuring that the thermometer on the pellet can obtain the temperature of the sample more accurately.

The choice of silver is due to three key factors. First, it has an excellent thermal conductivity at ultra-low temperature. Second, under the initial condition of the MCE measurements, the lack of demagnetization cooling characteristics in silver ensures that it does not interfere with the magnetocaloric effects of the samples. Third, the specific heat of silver is notably low ($<$ 0.1 mJ/mol/K or 9.26 $\times$ 10$^{-4}$ mJ/g/K at 100 mK) in the sub-Kelvin range~\cite{pobell2007}, significantly lower than that of the sample ($\sim$ 1 J/mol/K or 2.31 mJ/g/K at 100 mK), thereby exerting minimal impact on the lowest temperature of the entire sample pellet.

As a conductor, silver inevitably suffers from eddy current heating with varying magnetic field. The temperature rise induced by eddy current heating can be estimated based on a simple model with silver only~\cite{jang2015}. We take here the resistivity of silver $\rho$$\rm _{Ag}$ = 0.003 $\mu \Omega \cdot cm$ (residual resistivity ratio RRR = 500)~\cite{wu1995}, with the field ramp rate of $\frac{\partial B_0}{\partial t}$ = 50 Oe/s and the power loss per unit volume $P = \frac{R^2(\frac{\partial B_0}{\partial t})^2 }{8\rho}$ = 2.3 $\times$ $10^{-10}$ W. The sweeping of magnetic field from 6 T to 0 T took 20 minutes, during which the heat produced is $Q = Pt$ = 2.7 $\times$  $10^{-7}$ J. For a sample with the mass of $\sim$ 0.75 g and the specific heat of $C_{\rm sample}$ = 2.25 mJ/K, the temperature rise induced by the eddy current heating is $\Delta$$T$ = $\frac{Q}{C_{\rm sample}}$ = 0.13 mK, which is definitely negligible within the temperature scale of hundreds of millikelvins. In addition, once silver powders and the sample are homogeneously mixed into a pellet, the electrical conductivity is notably lower than that of pure silver, so that the actual eddy current heating will be even lower than the value estimated above. Consequently, the addition of silver powder does not have any negative effect on the results of MCE measurements.

\end{appendices}

\bibliographystyle{apsrev4-1}
\bibliography{NAMPref}


\end{document}